# Collision Prediction and Prevention in Contact Sports Using RFID tags and Haptic Feedback

Moeen Mostafavi[1], Fateme Nikseresht[1], Jacob Earl Resch[2],
Laura Barnes[1] and Mehdi Boukhechba[1]

[1]Department of Engineering Systems and Environment, [2]Department of Kinesiology,
University of Virginia, Charlottesville, VA, USA
{moeen, fn5an, jer6x, lb3dp, mob3f}@virginia.edu

**Abstract.** American football is a leading sport for contact-related injuries such as cervical spine injuries, some of which result from an unforeseen hit. The use of a feedback mechanism to alert an athlete of a potential hit may mitigate the risk for sport-related concussion and catastrophic injury by allowing athletes to effectively brace for an otherwise unforeseen impact. In this project, we created a proximity sensor system using radio frequency identification (RFID) technology to send haptic feedback to the wearer before a potential impact. Our pilot test consisted of three participants who wore our novel sensor system, who ran simulated routes to represent player-to-player contact. Additionally, we used simulated data from a Network Live Football (NFL) league game to further validate our contact prediction algorithm. Results show that our method can predict players' collisions with less than 14% false-alarms.

## 1 Introduction

Although the risk of impending injury in contact sports is high, those spots are prevalent. For example, football and basketball are the most popular sports in the United States. Over 1 million high school athletes and 70,000 college athletes play American Football. In these popular games, unanticipated hits in the game cause paralysis and other extreme injuries because athletes are unprepared for. The use of a feedback mechanism to alert an athlete of a potential hit may mitigate the risk for sport-related concussion and catastrophic injury by allowing athletes to brace for an otherwise unforeseen impact effectively.

   Concussion in contact sports is a constant concern for health professionals. Harmon et al. estimated that 3.8 million concussions occur in the US during competitive sports every year [8]. American football is a leading sport for contact-related injuries such as cervical spine injuries, resulting from an unforeseen hit that may range from 12g to 130g. Researchers measured the magnitude of linear (g) and rotational forces (rad/s) using accelerometers to evaluate the severity of concussive injuries during the games [1]. However, there is a debate on the accelerometer's accuracy in measuring the severity of head impacts [2]. Still, this technology helps researchers investigate how different athletes respond to the strike to their neck and head [3-5]. It can also help researchers to find the effect of different helmets and practices on head impacts [6,7].

A big challenge of using accelerometers is to minimize concussive injuries is their limitation to analyze the hit after the occurrence. To fill this gap in the literature, this project introduces a proximity sensor system using radio frequency identification (RFID) to predict incidents and send haptic feedback to athletes before a potential impact. Our pilot test consisted of three participants who wore our novel sensor system, who ran simulated routes to represent player-to-player contact. Additionally, we used simulated data from an NFL game as a proof-of-concept.

## 2   Methods

Our proposed Hit-Alert Optic Sensor is designed to predict and prevent collisions. For this purpose, we track players on the field using RFID sensors and send them haptic feedback to prevent injuries. This system consists of (a) a tracking system, (b) haptic feedback, and (c) the central server to detect impending severe collisions. In this section, we discuss how each element is implemented.

**(a) Tracking system:** Our hardware is based on WISER Redundant Radio Localization and Tracking (RRLT) System which has a set of RFID tags that can be placed anywhere on the football players bodies. The players carry tracking tags and the system provides up to 12Hz sampling rate. These tags are tracked in real-time using a set of five antennas to track their position.

In this project, we installed four antennas in the corner of the game field, and the fifth one, which was connected to the server, was located along the half court line. Fig. 1 shows how these hardware tools look and how we can install them on a football court.

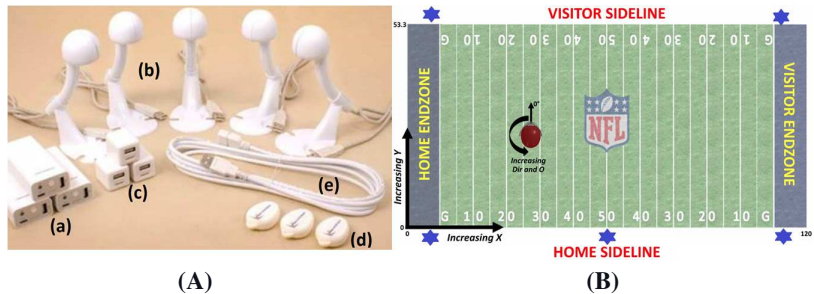

**(A)**                                                                                    **(B)**

**Fig. 1.** Player real time tracking system using RFID technology. (A) The system hardware includes (a) batteries for the antenna, (b) antennas to track the tags, (c) chargers, (d) RFID tags that are carried by the players, and (e) USB extension cables. The antennas can be installed on the corners of the field and the fifth one can be placed anywhere between any two corners. In figure (B) a recommended place to install the antennas are shown with blue stars.

The antennas can only track the location of RFID tags. Installing two tags on the players' shoulder pads help us to find their orientations during the game.

**(b) Haptic feedback:** For the haptic feedback, we used a TX-7470-C232 interface transmitter from the LRS paging system. This device provides a vibration stimulus to the torso of players who are projected to be in a severe collision. Every time an incident was anticipated, the player received a vibration, and blinking LEDs of the pager made it visible to other people.

**(c) Central server:** The fifth antenna of the RFID system connects to the server computer using a USB port. The data is transferred to the local host in json format. The TX-7470-C232 interface uses COM Port to transfer the data using the terminal. To process the data, Matlab imports the data from the local host then analyzes in real time to send feedback. For future analysis, the data is stored on the server computer and the pilot test was video captured.

## 3     Results

We had three phases to test the proposed approach. First we collected data and processed it offline for initial tests. We compared different estimation approaches in this phase. After deciding on implementation details, we implemented an online system in a basketball practice facility. Finally, we simulated data from a real football game.

### 3.1     Initial tests

Initially we implemented the system in a large office and the server program only stored the data without any processing. We processed the data offline in Matlab. The GUI platform shown in Fig. 2 used to simulate the tracking system under different scenarios.

After implementing different scenarios in Matlab, we decided on the following issues:

(a) The location data is noisy, and we can see several jumps for each sensor. Also, in many steps the data for some of the tags are missed. To address this problem, we used two tags for one player and their average values are considered as the location for the player. Still the level of noise was high, so we used a weighted average of the last three samples of each sensor.

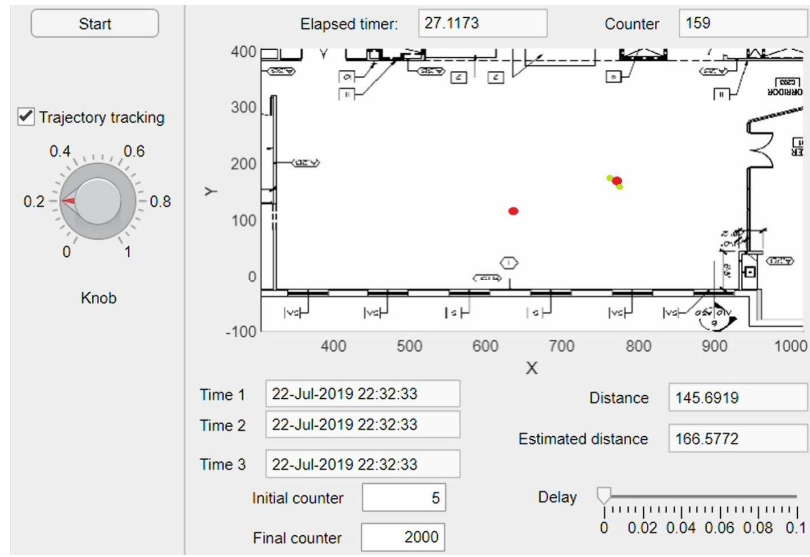

**Fig. 2.** Simulating tracking data in Matlab. red circles show the location of participants in the office. The timestamp for the most recent data sample is shown at the bottom left side of the map. The delay slider changes simulation speed, and the knob gets parameter values from the user. On the right bottom side of the map, the distance of participants and its estimated value for the next step is shown.

(b) The actual sampling rate of the sensors after activating the noise filters was about 5 samples per second. In this case we cannot track a person running very fast. For the tests we asked the participants to walk fast instead of running. Since this problem can be resolved using sensors with higher sampling rate, this is not a limitation for generalizing the result.

(c) The processing, visualization, and communication time are not small and their delay is significant. We decided to remove the visualization from real-time implementation and made the processing algorithm simple.

(d) RFID tags give us only locations but in ideal cases we have speed and acceleration to predict the location in the next steps. Even if acceleration and speed are known in the beginning, they are not fixed during a real game. We used NFL data and tested different scenarios to find which one has a good accuracy and minimum false alarm. We found assuming fixed speed between two samples gives the best result.

(e) The incidents are defined when the distance between two players is decreasing below a given threshold. In a real game we can consider a threshold of three feet, however, due to the participant's speed and sampling rate limitation we cannot consider the same threshold in our test. We found a threshold of two feet gives us a reasonable result.

### 3.2 Implementation

To test our framework, four antennae were installed in the corners of an indoor basketball practice facility. The fifth antenna was connected to the server computer and located along the half-court line. In the pilot study, three participants carried the RFID tags in their pockets. The haptic feedback device was affixed to each participant's chest at the level of the manubrium. Participants walked and jogged routes to simulate football routes with several planned events representing player-to-player contact.

Incidents are expected when the distance between two players is decreasing, and they are less than 2 feet apart. RFID data was monitored using Matlab in real-time to send feedback. Whenever an incident was anticipated, the player received a vibration via the haptic device affixed to their chest. Routes were recorded via a digital camera. The predicted collision was identifiable on the cameras via blinking LEDs of the haptic device. Fig. 3 shows its implementations in the field.

In this implementation players did not move very fast and the system could predict incidents between the players. As we can see in Fig. 3, the haptic feedback is sent when the players get close to each other. This implementation is video captured and can be shared upon request.

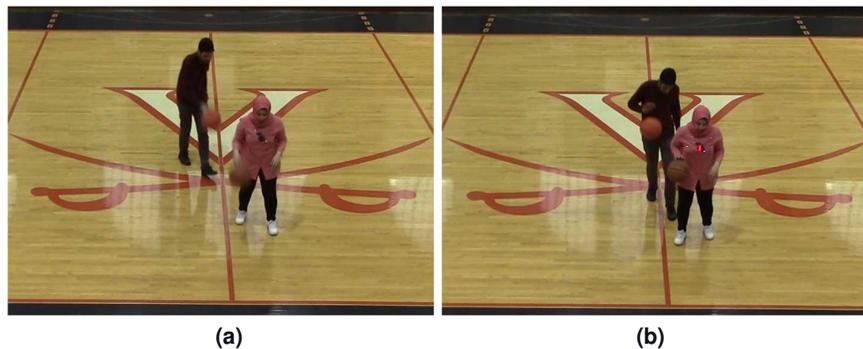

(a)          (b)

**Fig. 3.** Implementing in a basketball practice facility. (a) One player is approaching the other one from the behind and their distance is more than 3 feet. (b) The distance between players is going less than three feet and the red blinking LEDs show the player is receiving the haptic feedback.

### 3.3 Simulation

Zebra Technologies has been working with the National Football League (NFL) to provide player-tracking technology. NFL has installed two RFID tags in the player's shoulder pads to collect tracking-data. In 2018, the NFL started a contest called Big Data Bowl and released tracking data for 2017 games. The data is publicly available [9].

To further validate our collision detection algorithm, we used data generated from the NFL. We used this data for simulating our algorithm in a real game. In this simulation, athletes would have circles appear around the area that collisions are anticipated and actually occurred. Fig. 4 one predicted incident and screenshot from the real game.

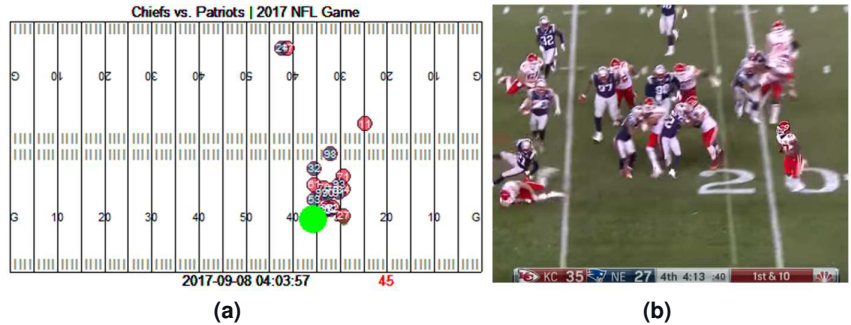

**Fig. 4.** 2017 NFL game between Chiefs vs. Patriots (a) Simulated games using collected RFID data by Zebra company. Predicted incident location is highlighted in green. (b) Youtube video capture of the incident

NFL data includes the location, speed, and orientation of the players in the game. But it is not clear how the orientation and speed of the players are found from RFID data. Since the data is released after the game, they may have used the location of the players in the next steps to find the speed and orientation. In that case, these metrics can not be used in a real-time implementation. For this reason, we considered location estimation in two different approaches: (a) the speed and orientation of the players are given in real-time, (b) the players have constant speeds between two samples (0.1 seconds), and we estimate the speed and location ourselves. The real data and the two approaches in a less than 15-seconds game simulation are shown in Table 1.

Table 1. includes the data from a real game, and there are six major incidents when the threshold is two feet. Considering a constant speed between two samples, we can predict all these incidents plus a false alarm for an incident in 28th frames. For the first actual incident, the fixed speed results in a 0.4-second error for the incident time, but all the other five incidents are accurately estimated. So in this approach, we have only 14% false-alarm. Using given speed and orientation from NFL data, we have four false-alarms, and two incidents are missed.

**Table 1.** Predicted incidents vs. actual ones. when the incident is defined in cases that players have a distance of less than two feet. The players that are involved in the incident and the frame that incident occurred are shown in this table.

| Index | Player A | Player B | Real data | Constant Speed | NFL Speed |
|---|---|---|---|---|---|
| 1 | (61) Mitch Morse | (97) Alan Branch | | 28 | 34 |
| 2 | (88) Ross Travis | (52) Elandon Roberts | | | 35 |
| 3 | (84) Demetrius Harris | (91) Deatrich Wise | 36 | 32 | 36 |
| 4 | (72) Eric Fisher | (52) Elandon Roberts | 41 | 41 | |
| 5 | Travis Kelce | (21) Malcolm Butler | 45 | 45 | |
| 6 | (72) Eric Fisher | (52) Elandon Roberts | 53 | 53 | 52 |
| 7 | (88) Ross Travis | (52) Elandon Roberts | 54 | 54 | 53 |
| 8 | (72) Eric Fisher | (88) Ross Travis | 62 | 62 | 61 |
| 9 | (21) Malcolm Butler | (90) Malcom Brown | | | 62 |
| 10 | (21) Malcolm Butler | (91) Deatrich Wise | | | 71 |

## 4 Concluding Remarks

Fatality and serious injuries in contact sports is a very concerning factor for athletes. This study introduced a novel framework to predict concussion injuries during a game and send haptic feedback to minimize the risk of these incidents.

NFL players run more than 22 mph, which means they are going more than 1-yard in 0.1 seconds. Since the RFID sensors' sampling rate was 10Hz, some incidents with thresholds less than 1 yard are missed. Having a sampling rate higher than 15 Hz can substantially improve the result.

We used a commercial paging system for haptic feedback, which is usually used for paging every couple of seconds. For future work, the haptic feedback system should be designed for this application.

Implementing this system increases safety for the players and provides data for further analysis by the coaches. The low cost of RFID tags and pagers makes the system affordable for high school and college teams.